\documentclass[aps,prl,twocolumn,showpacs,superscriptaddress]{revtex4-1}

\usepackage{graphicx}
\usepackage{amsfonts, amsmath, amstext, amssymb, amsfonts, amsxtra}
\usepackage{braket}
\usepackage{bbm}
\usepackage{bbold}
\usepackage[protrusion=true,expansion=true]{microtype}
\usepackage{wasysym}
\usepackage{cleveref}
\usepackage{color}
\usepackage{ulem}

\DeclareMathOperator{\tr}{tr} 
\newcommand{\im}{{i}}         
\newcommand{\St}{\mathcal{S}} 



\begin{document}
	
	\title{Quantum Lyapunov exponents beyond continuous measurements}
	\author{I. I.~Yusipov$^1$, O. S.~Vershinina$^1$, S.~Denisov$^2$, S. P.~Kuznetsov$^3$, and M. V.~Ivanchenko$^1$}	
	\affiliation{ $^1$ Department of Applied Mathematics, Lobachevsky University, 603950, Nizhny Novgorod, Russia\\
		$^2$ Department of Computer Science, Oslo Metropolitan University, N-0130, Oslo, Norway, \\
		$^3$ Kotel’nikov’s Institute of Radio-Engineering and Electronics of RAS, 410019, Saratov, Russia}
	
	\begin{abstract} 
		Quantum systems, when interacting with their environments, may 
		exhibit  non-equilibrium states that are  tempting to be interpreted as quantum analogs of
		chaotic attractors. However, different from the Hamiltonian case, 
		the toolbox for quantifying dissipative quantum chaos remains  limited. In particular, quantum generalizations of Lyapunov exponents, the main quantifiers of classical chaos, 
		are established only within the framework of continuous measurements. 
		We propose an alternative generalization based on the unraveling of quantum master equation into an ensemble of `quantum  trajectories', 
		by using the so-called  Monte Carlo wave-function method. 
		We illustrate the idea with  a periodically modulated open quantum dimer and demonstrate that the transition to quantum chaos matches the period-doubling 
		route to chaos in the corresponding mean-field system.           
	\end{abstract}

	\maketitle
	
	\textbf{It is one of the pillar concepts of Chaos theory that complex deterministic dynamics is rooted in the local
 instability  which forces two initially close trajectories to diverge. 
	This divergence 
	is conventionally quantified with Lyapunov exponents (LEs), a powerful tool to quantify  dynamical chaos. The history of attempts to generalize LEs to 
	quantum dynamics is nearly as old as the history of Quantum Chaos. 	Most of this history is about the Hamiltonian limit, where the spectral 
	theory of Quantum Chaos \cite{Haake1991}  was  established first. The corresponding generalizations range from early ideas to use quasi-probability functions 
	and define quantum LEs in terms of a ``distance'' between them \cite{Toda1987,Karol1992,Manko2000} to very recent advances based on out-of-time correlation 
	functions \cite{Galitski2017,Galitski2018,Hirsch2019}. When a quantum system is open and its dynamics is modeled with a quantum master equation \cite{book}, 
	the evolution of the system's density operator can be unraveled into an ensemble of evolving  trajectories, each one described by a wave function \cite{book}. 
	Dynamics of these wave functions is essentially stochastic; therefore, LEs could be  introduced in a more intuitive way 	than in the Hamiltonian limit. 
	But will so-defined exponents make sense? Here we define a particular type of quantum  LEs and give a positive answer to this question.  Since quantum 
	trajectories \cite{Carmichael1991} are not just a formal trick but a part of reality, e.g., in optical \cite{Walter2006} and microwave \cite{Nori2017} cavity systems, 
	we believe that our results will be of interest to the theoreticians (and, hopefully, to the experimentalists) dealing with these systems.
	}
	
	\section{Introduction}
	Hamiltonian chaos, a fascinating product of the sensitivity to initial conditions and topological mixing co-working in the phase space of classical nonlinear systems, has been extended to the quantum realm quite early. 
	As a result, a profound understanding of the spectral signatures of Hamiltonian quantum chaos \cite{Casati1979,Gutzwiler1991,Haake1991,Guhr1998} has been reached. Quantum generalization of Lyapunov exponents, one of the main quantifiers of the Hamiltonian classical chaos, has also been at the focus of intensive studies during last three decades \cite{Toda1987,Haake1992,Aleiner1997,Kuznetsov1998, Kuznetsov2000}; this topic experiences now yet another revival, see Refs.~\cite{Maldacena2016,Galitski2017,Galitski2018,Hirsch2019}.

	The fast progress in experimental quantum physics, especially in such fields as cavity quantum electrodynamics \cite{Walter2006}, quantum optical systems \cite{Aspelmeyer2014}, 
	artificial atoms \cite{You2011} and polaritonic devices \cite{Feurer2003}, has diverted attention from the ideal Hamiltonian limit to a more realistic description. 	
	All the corresponding systems are open, i.e., they interact with their environments (or are subjected to actions from outside), and therefore their dynamics is essentially 
	dissipative \cite{book,carm}. In turned out that this type of quantum evolution is no less complex and versatile than the unitary one \cite{Diehl2008,Budich2015}.  
	
	There is ample evidence, both  computational and experimental, that  asymptotic states of open far-out-of-equilibrium quantum systems 
	can yield (when measured, e.g., by means of quantum tomography)  structures similar to classical
	chaotic attractors \cite{Spiller1994,Brun1996,Hartmann2017,Ivanchenko2017,Poletti2017,Poletti2018}. 	
	However, quantification of dissipative quantum chaos remains little explored. Approaches attempting to 
	match variations in the spectra of generators of dissipate quantum evolution \cite{Grobe1988} or their zero-eigenvalue 
	elements (asymptotic density operators) \cite{Prosen2013,Hartmann2017,Ivanchenko2017} with  transitions between regular and chaotic 
	regimes in the corresponding mean-field equations, have brought some interesting results. However, at the moment, these findings are supplemented only by conjectures 
	and speculations. 
	
	How to generalize LEs, or, for a start,
	{\it the largest} LE, to the case of open quantum systems? 	A promising strategy is (1) to unravel the solution of the master equation, governing 
	the evolution of the systems density operator, into a set of quantum trajectories and then (2) to quantify (somehow) the divergence rate between initially  close trajectories.

	Strictly speaking, there are infinitely many ways to unravel a given master equation but very few of them are physically plausible \cite{Wiseman1992}. To the best of our knowledge, the only existing  realization of this idea is related to the framework of diffusive-type continuous measurements, which deals with trajectories of the stochastic Schr\"{o}dinger equation \cite{Persival1992}. 
	This approach allowed, e.g., to obtain Lyapunov exponents, based on the expectation values of $x$ and $p$ observables,
	for the quantum version of periodically modulated Duffing oscillator, albeit in the vicinity of the classical limit \cite{Jacobs2000,Pattanayak2018, Eastman2018}. 
	It allows (potentially) to go deeper into the  quantum regime; however,  the decreasing value of Lyapunov exponent makes discrimination between regular and chaotic quantum states a hard task \cite{Ota2005,Pattanayak2008,Jacobs2009,Pattanayak2009}.        
	
	The continuous measurement framework has a perfect physical meaning.
	For example, it describes an optical cavity whose output is monitored via homodyne detection \cite{Carmichael1991,Wiseman1992}; this
	creates a perspective to measure LEs in an experiment \cite{Eastman2018}. However,  high computational cost  
	compromises the physics-related benefits of this approach and
	constrains numerical studies to the resolution of single-parameter behavior of model systems \cite{Jacobs2006,Pattanayak2018}.      
	
	In this paper we propose an alternative approach to quantum generalization of Lyapunov exponents. It defines
	the largest quantum Lyapunov exponent in terms of the
	`quantum jump' unraveling based on the  Monte Carlo wave-function (MCwf) method  \cite{zoller,dali,plenio,daley}.
	This unraveling is very relevant in the context of quantum optics and cavity systems \cite{Carmichael1991}.  	Computational 
	efficiency of the currently available numerical implementations of the MCwf method \cite{Volokitin2017} has allowed us to explore the  parameter 
	space of a scalable quantum model and reveal a complex structure of intermingled regular and chaotic domains. 
	We also go deep into the quantum regime and quantify there transitions corresponding to the classical period-doubling route to chaos.  
	
	\section{Model}
	Within the Markovian approximation framework (which assumes weak coupling to environment), the evolution of an open quantum system can be described by the Lindblad master equation \cite{book, alicki},
	\begin{align}
	\dot{\varrho} = \mathcal{L}(\varrho) = -\im [H,\varrho] + \mathcal{D}(\varrho),
	\label{eq:1}
	\end{align}
	where the first term in the r.h.s.\ captures the unitary evolution, 
	and the second term describes the action of environment. We consider a system of $N$ indistinguishable interacting bosons, 
	that hop between the sites of a periodically rocked dimer. This model is  a popular theoretical testbed  \cite{Vardi, Witthaut, PolettiKollath2012}, 
	recently implemented in experiments \cite{oberthaler, ober1}, 
	known to exhibit regular and chaotic regimes \cite{Hartmann2017,Ivanchenko2017,Poletti2017,Poletti2018}. Its unitary dynamics is governed by the Hamiltonian 
	\begin{align}
	H(t)=&-J \left( b_1^{\dagger} b_2 + b_2^{\dagger} b_1 \right) + \frac{2U}{N} \sum_{g=1,2} n_g\left(n_g-1\right)\nonumber \\
	&+\varepsilon(t)\left(n_2 - n_1\right) \;. 
	\label{eq:2}
	\end{align}
	Here, $J$ denotes the tunneling amplitude, $U$ is the interaction strength, and $\varepsilon(t)$ presents a periodical modulation of the on-site potentials. 
	In particular, we choose
	$\varepsilon(t)=\varepsilon(t+T)=\mu_0+\mu_1 Q(t)$, where $\mu_0$ and $\mu_1$ denote static and dynamical energy offsets between the two sites, respectively. $Q(t)$ itself is a periodic unbiased two-valued quench-function with one full period $T$; more specifically, $Q(\tau) = 1$ within $0 < \tau \le T/2$
	and $ Q(\tau) = -1$ for the second half period $ T/2 < \tau \leq T$.
	$b_g$ and $b_g^{\dagger}$ are the annihilation and creation  operators on sites $g \in \{1,2\}$, while $n_g=b_g^{\dagger}b_g$ is the particle  number operator. 
	The system Hilbert space has dimension $N+1$ and can be spanned  with $N+1$ Fock basis vectors, labeled by the number of bosons on the first  site $n$, $\{|n+1\rangle\}$, $n=0,...,N$. Thus, the size of the model is controlled by the total number of bosons.
	
	The dissipative term involves a single jump operator \cite{Diehl2008}:
	\begin{align}
	\mathcal{D}(\varrho) =& \frac{\gamma}{N} \left(V\varrho V^\dagger - \frac{1}{2}\{V^\dagger V,\varrho\}\right), \\
	V=&(b_1^{\dagger} + b_2^{\dagger})(b_1-b_2), 
	\label{eq:3}
	\end{align}
	which attempts to `synchronize' the dynamics on the two sites by constantly recycling anti-symmetric out-phase modes into symmetric in-phase ones. The dissipative coupling constant $\gamma$ is taken  to be
	time-independent.  Throughout the paper we will assume $J=1, \mu_0=1, \gamma=0.1$ and $T=2\pi$.
	
	Now we employ the MCwf method \cite{zoller,dali}	to unravel  deterministic  equation (\ref{eq:1}) into an ensemble of quantum trajectories. 	
	It recasts the evolution of the model system into evolution of the ensemble of systems described by wave functions, $\psi_r(t)$, $r = 1,2,...,M_r$,  
	governed by an effective non-Hermitian Hamiltonian, $\tilde{H}$. This Hamiltonian incorporates the dissipative operator $V$, which is responsible for the decay of the norm,
	\begin{align}
	i\dot{\psi}=\tilde{H}\psi, \ \tilde{H} = H -\frac{i}{2} V^\dagger V.
	\label{eq:4}
	\end{align} 
	When the norm drops below a randomly chosen threshold, 
	the wave function is transformed according to $\psi\rightarrow V \psi$ and then normalized \cite{Carmichael1991}.

	\begin{figure}[t]
		\begin{center}
			\includegraphics[width=0.95\columnwidth,keepaspectratio,clip]{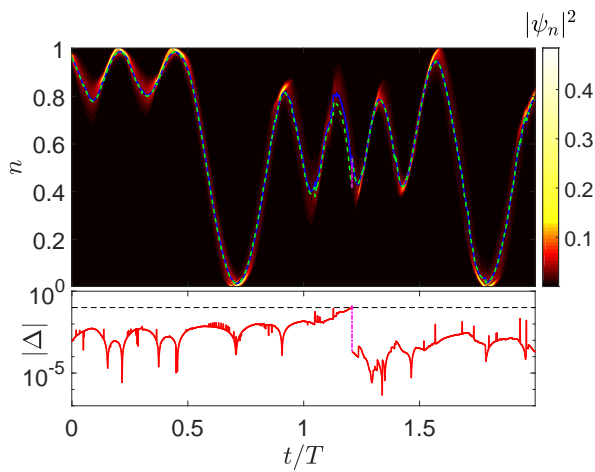}
			\caption{(Color online) Evolution of fiducial and perturbed trajectories. Top panel: expectations $n_f(t)$ (blue, solid) and $n_a$ 
			(green, dashed), together with fiducial wave function amplitude $\psi_f(t)$ (color coded). Bottom panel: evolution of the mismatch between 
			the expectations for the two quantum trajectories, $\Delta(t)$. Visible spikes correspond to single quantum jumps, magenta vertical line indicates 
			resetting of the perturbed trajectory when the mismatch goes above the threshold $\Delta_{max}=0.1$ (black dashed horizontal line). Other parameters are $U=0.5, \mu_1=1.5$, and 
			$N=200$.  } 
			\label{fig:1}
		\end{center}
	\end{figure}

	The density matrix can then be sampled from a set of $M_r$ realizations as
	$\varrho(t_\mathrm{p};M_{\mathrm{r}}) = \frac{1}{M_{r}}
	\sum_{j=1}^{M_{\mathrm{r}}} \ket{\psi_j(t_\mathrm{p})}\bra{\psi_j(t_\mathrm{p})}$, which, given an initial pure state $\psi^\mathrm{init}$, 
	converges towards the solution of Eq.~(\ref{eq:1}) at time $t_\mathrm{p}$ for the initial density matrix $\varrho^{\mathrm{init}} = \ket{\psi^\mathrm{init}}\bra{\psi^\mathrm{init}}$. 
	We make use of the recently developed high-performance realization of 
	the quantum jumps method \cite{Volokitin2017} and generate $M_{\mathrm{r}}=10^2$ different trajectories for averaging, 
	leaving $t_0 = 2\cdot10^3 T$ time for relaxation towards an asymptotic state, and following the dynamics for up to $t = 10^3T$.

	The further analysis is focused on the expectation values of two observables, these are a normalized number of particles on the left site of the dimer, $n(t)$, and the energy, $E(t)$,  
	\begin{align}
	n(t)&=\frac{1}{N} \langle\psi(t)|b^\dagger_1b_1|\psi(t)\rangle, \\
	E(t)&=\langle\psi(t)|H|\psi(t)\rangle.
	\label{eq:5}
	\end{align}
	The former observable has a phase variable counterpart in the nonlinear mean-field equation for the classical model that will be introduced later.

	\section{Definition of the largest Lyapunov exponent}
	
	The largest quantum Lyapunov exponent  is calculated as the average rate of the exponential growth of the ``distance'' (defined with some metrics)  between the fiducial and auxiliary trajectories, $\psi_f(t)$ and $\psi_a(t)$, evolving under Eq.~(\ref{eq:4}), in full analogy to the classical definition \cite{Benettin1976}.

	In our approach, the distance is defined as the 
	difference between expectation values -- along the fiducial and auxiliary trajectories -- of some, preliminary chosen,  operator.  
	The auxiliary trajectory is initialized 
	as a normalized perturbed vector $\psi_a^{init}=\psi_f^{init}+\varepsilon\psi_r$, produced with random i.i.d. entries in $\psi_r$ and $\varepsilon\ll1$, 
	adjusted so that the initial difference between the fiducial and perturbed observables, $n_f(t)$ and $n_a(t)$, $\Delta_0=|n_f(0)-n_a(0)|$ is equal to a certain fixed value. 
	Fig.~\ref{fig:1} shows that the wave function $\psi(t)$ remains well-localized during evolution even in the aperiodic regime, 
	and the fiducial and perturbed observables remain close to each other after many quantum jump events. As the difference $\Delta(t_k)=|n_f(t)-n_a(t)|>\Delta_{max}$ 
	exceeds the threshold at $t=t_k$, the perturbed state is renormalized close to the fiducial one along the mismatch direction $\psi_f(t)-\psi_a(t)$, so that it returns $|n_f(t)-n_a(t)|=\Delta_0$, 
	and the growth factor $d_k=\Delta(t_k)/\Delta_0$ is recorded \cite{Benettin1976}. Finally, the largest LE  is  estimated as 
	\begin{equation}
	\label{eq:7a}
	\lambda=\lim\limits_{t\rightarrow\infty}\frac{1}{t}\sum\limits_k\ln d_k.
	\end{equation}

\begin{figure}[t]
 	\begin{center}
 		\includegraphics[width=0.95\columnwidth,keepaspectratio,clip]{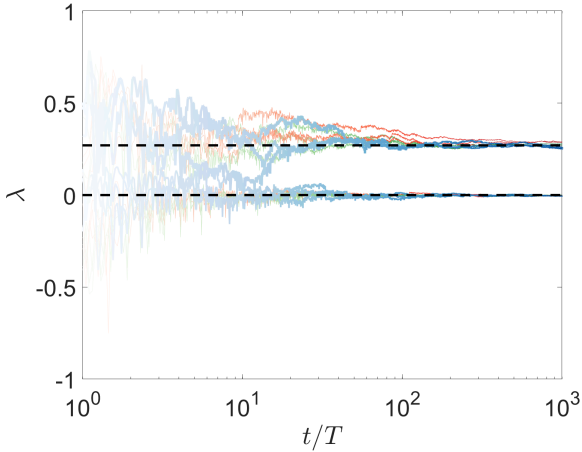}
 		\caption{(Color online) Finite time Lyapunov exponents converging to $\lambda=0$ (regular) 
 		and $\lambda\approx0.27$ (chaotic regime) for three individual trajectories for a given set of parameters (thin lines). 
 		We use as observable $n(t)$ [$U=0.05$ and $U=0.5$ (green)] and $E(t)$ [$U=0.05$ and $U=0.5$ (red)]. 
 		The parameters are  $\Delta_0=10^{-4}$ and $\Delta_{max}=0.1$. 
 		Thick lines correspond to the  exponent calculated by performing renormalization after fixed time, $t_k=kT$, for observable $n(t)$, $U=0.05$ and $U=0.5$ (light blue).  
 		 		 Line color saturates to maximal value upon increase of time. Other parameters are $\mu_1=1.5$ and $N=200$.}  
 		\label{fig:2}
 	\end{center}
 \end{figure}

	Our definition is inspired by the concept  of LEs introduced for  classical jump systems \cite{Fang}. These are linear systems 
	whose evolution is interrupted with jumps events, which themselves are governed by a finite-state Markov process. There LEs are calculated at the jump instances and, therefore,
	renormalization are subordinated to the sojourn times of the underlying Markov process. In fact, the value of the LE is independent of the particular distribution of 
	the times when the normalization is performed -- provided that the mean time (first moment of the distribution) is fixed. 
	For example, the exponent can be calculated by performing the renormalization
	after fixed time so that  $t_k=k\tau$, as in Ref.\cite{Benettin1976,Pikovsky_Politi}. If the $\tau$ is matching the mean time between the crossing then two LEs converge to  close values, 
	see Fig.~\ref{fig:2}.

\begin{figure}[t]
		\begin{center}
			(a) \includegraphics[width=0.9\columnwidth,keepaspectratio,clip]{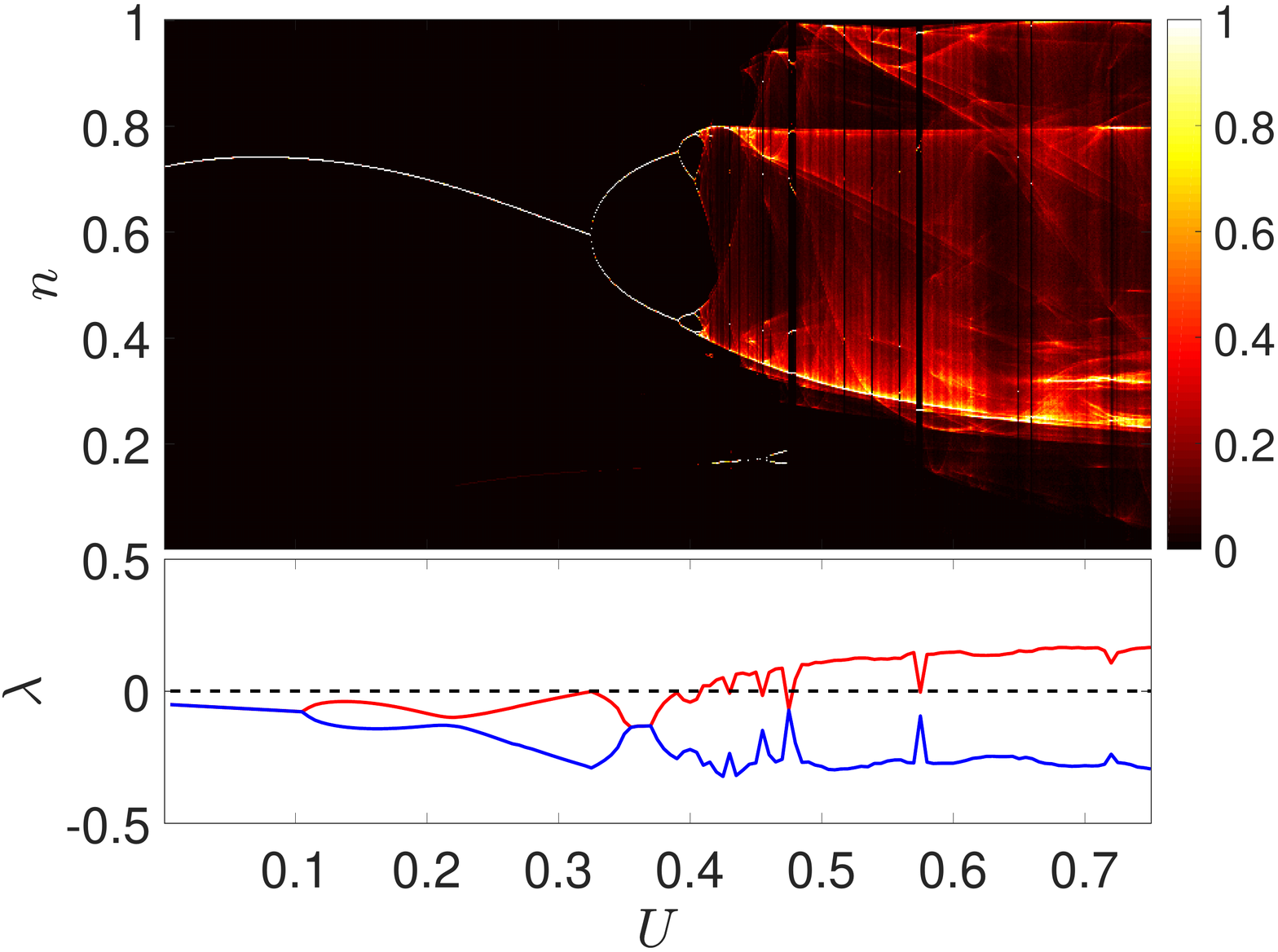}\\
			(b) \includegraphics[width=0.9\columnwidth,keepaspectratio,clip]{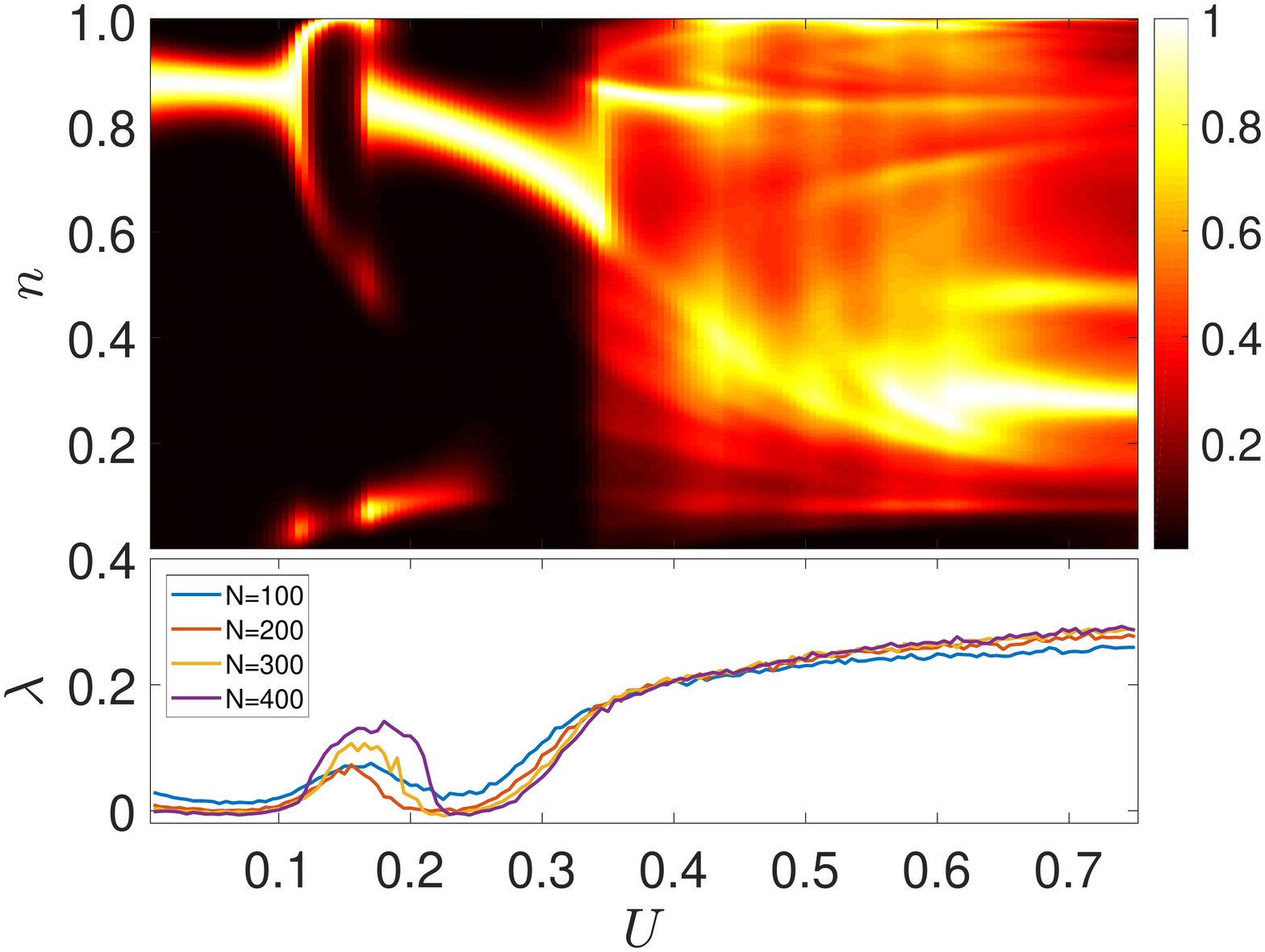}
			\caption{(Color online) Classical-quantum correspondence in the chaos development and
			Lyapunov spectrum as  function of the interaction strength $U$. (a) Mean-field equations:  color-coded bifurcation histogram for the  
			stroboscopic map[top panel] and  Lyapunov spectrum [bottom panel]. (b) Quantum model: color-coded probability to observe a 
			fraction of particles $n$ in the asymptotic regime (the maximal element normalized to $1$) [top panel] and  quantum largest 
			Lyapunov exponent for different number of particles $N$ [bottom panel]. The parameters are  $J=1, \mu_0=1, \mu_1=1.5$, and $N=200$.} 
			\label{fig:3}
		\end{center}
	\end{figure}

	To check our idea we consider two  interaction strength values, $U=0.05$ and $U=0.5$ (the other parameters are $\mu_1=1.5, N=200$), 
	for which regular and chaotic regimes in mean-field equations have been previously identified \cite{Ivanchenko2017}. 
	Lyapunov exponents calculated by the means of individual quantum trajectories converge to their asymptotic values upon increase of $t$; see Fig.~\ref{fig:2} (thin lines). 
	We also observe that for $N \geqslant 200$ the largest LE saturates to the size-independent asymptotic value (Fig.~\ref{fig:3}(b;bottom panel)). 
	
	The result has proved to be independent on the choice of a particular observable, $n(t)$ or $E(t)$; 
	both yield near the same asymptotic values, up to numerical resolution of the method (see the relevant discussion in the Conclusions).
	Further on, we use $n$ as an observable, and calculate the largest LE based on averaging over $M_r=100$ trajectories; see.  
	
	We also calculated the largest LEs by following the standard prescription\cite{Benettin1976}  and performing renormalization after fixed time $\tau = T$, see Fig.~\ref{fig:2} (thick lines). 
	There one only has to ensure that the mismatch $\Delta(t)$ remains small over $\tau$, when the largest LE is positive; for the chosen renormalization time $\tau$, 
	it requires  $\Delta_0 \leq 10^{-6}$. We also observed some weak variability in the limiting value of LE depending 
	on the parameters of the method, $\Delta_0, \Delta_{max}, \tau$, a property genuine to classical nonlinear systems \cite{Pikovsky_Politi,Cencini2013}.

	\section{RESULTS}
	
	Here we analyze the transition to dissipative quantum chaos. 
	First, we introduce the corresponding mean-field equations as a reference. In the limit $N \rightarrow \infty$, 
	the dynamics of quantum dimer can be approximated by the equations for expectation values of  three pseudo-spin operators $\St_x=\frac{1}{2N}\left(b^{\dagger}_1 b_2 + b^{\dagger}_2 b_1\right)$, $\St_y= -\frac{\im}{2N}\left(b^{\dagger}_1 b_2 - b^{\dagger}_2
	b_1\right)$, $\St_z=\frac{1}{2N}\left(n_1 - n_2\right)$. For a large number	of atoms, the commutator $\left[\St_x,\St_y\right]= [\im\St_z/N]
	{\overset{N\to\infty}{=}} 0$ and similarly for other cyclic permutations.
	Replacing operators with their expectation values, $\braket{\St_k} =\tr
	[\varrho \St_k]$, and denoting $\braket{\St_k}$ by $S_k$, one obtains the
	semi-classical equations of motion~\cite{Hartmann2017}
	\begin{align}
	\dot{S}_x &= 2\varepsilon(t)S_y - 8U S_zS_y + 8\gamma \left(S_y^2+S_z^2\right), \nonumber\\
	\dot{S}_y &= -2\varepsilon(t)S_x + 8U S_xS_z +2JS_z - 8\gamma S_xS_y, \nonumber\\
	\dot{S}_z &= 2JS_y - 8\gamma_0 S_xS_z.
	\label{eq:6}
	\end{align}
	As $S^2=S_x^2+S_y^2+S_z^2 = 1/4$ is a constant of motion, one
	can reduce the mean-field evolution to the surface of a Bloch sphere,
	$\left(S_x,S_y,S_z\right) = \frac{1}{2} (\cos\varphi\sin\vartheta,\sin\varphi\sin\vartheta, \cos\vartheta )$,
	yielding the equations of motion
	\begin{align}
	\dot{\varphi} &= 2J\frac{\cos\vartheta}{\sin\vartheta}\cos\varphi - 2\varepsilon(t) + 4U \cos\vartheta - 4\gamma \frac{\sin\varphi}{\sin\vartheta}, \nonumber \\
	\dot{\vartheta} &= 2J\sin\varphi + 4\gamma \cos\varphi\cos\vartheta \label{eq:7}\,.
	\end{align}
	A convenient choice to match the quantum and classical solutions is to follow the fraction of particles at the first site, which classical counterpart is $n(t)=[1+\cos\theta(t)]/2$. 
	
	Upon tuning parameter values, the nonlinear  mean-field equations display complex dynamics; in particular, they exhibit period-doubling route to chaos \cite{Hartmann2017,Ivanchenko2017,Poletti2018}. 
	Fig.~\ref{fig:3} shows a  bifurcation diagram, which depicts the marginal probability density function (pdf) of  stroboscopic values of $n$, as function of $U$. For each value of $U$ it was sampled with $10^4$ values, $n_k=n(t_0+kT)$, $k=1,2,...,10^4$, generated by the flow, Eq.(\ref{eq:7}), after the transient time $t_0 = 10^4T$.  One Lyapunov exponent becomes positive as the chaotic attractor emerges, another remains negative.   
	
	\begin{figure}[t]
		\begin{center}
			(a)\includegraphics[width=0.9\columnwidth,keepaspectratio,clip]{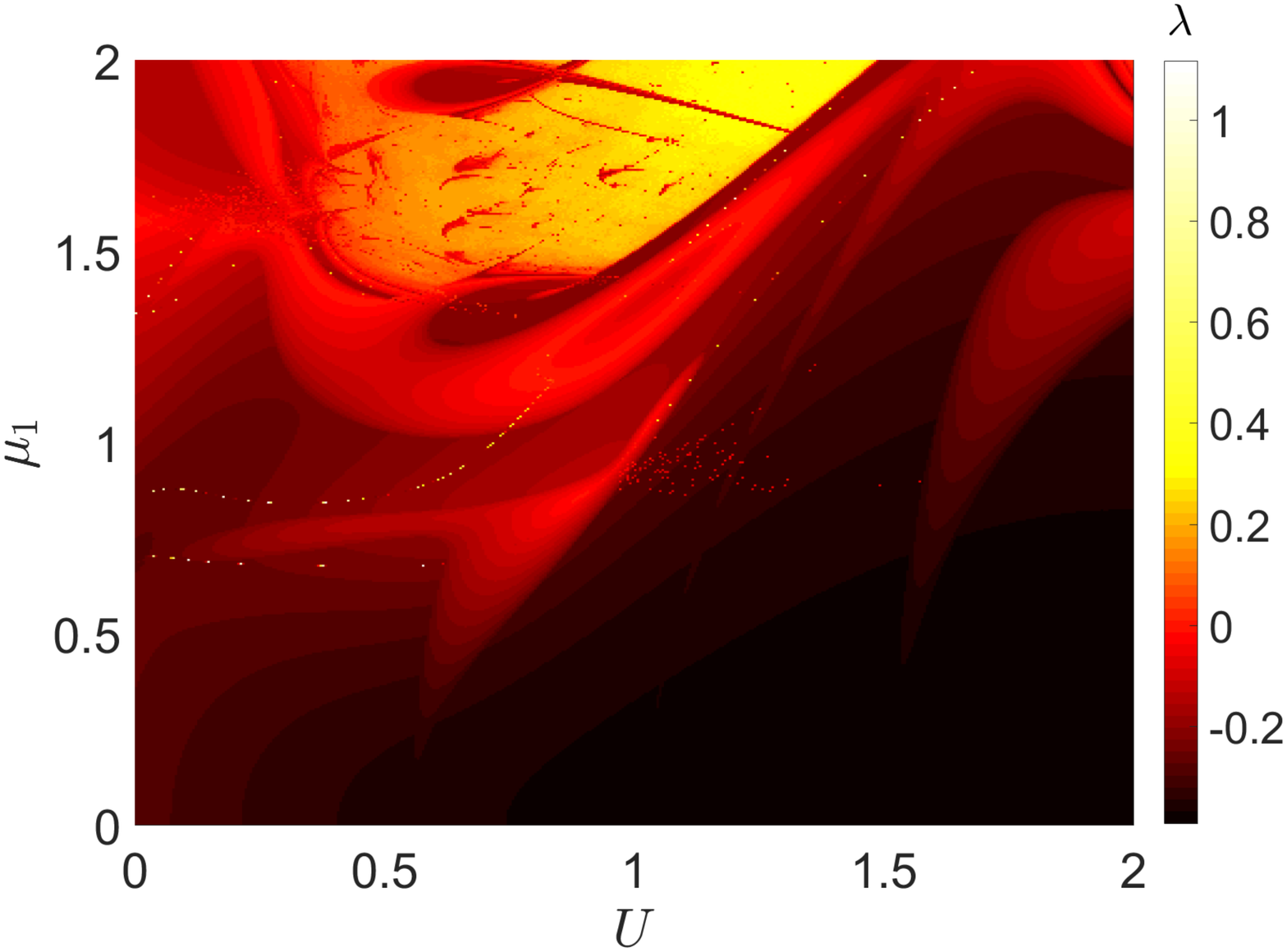}
			(b)\includegraphics[width=0.9\columnwidth,keepaspectratio,clip]{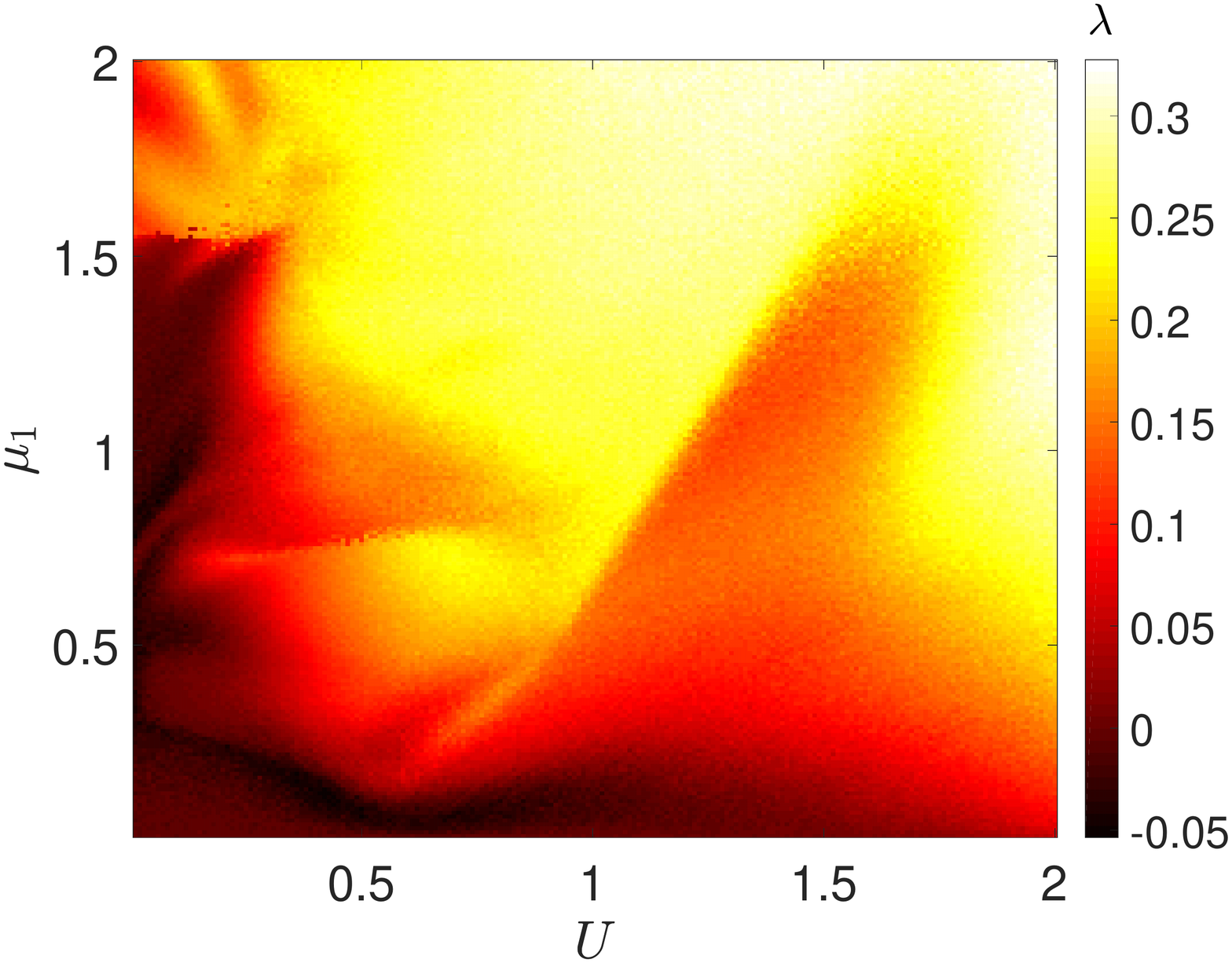}
			\caption{(Color online) Chaos-order phase diagram on the  parameter plane 'interaction strength $U$ -- driving 
			amplitude $\mu_1$' for the   mean-field equations, Eq.~(9), (a) and  open quantum dimer, Eqs.~(2-4), (b).  
			The color-coded quantum Lyapunov exponent indicates regular and chaotic regimes. The number of bosons is $N=100$; other parameters as in Fig.~3.} 
			\label{fig:4}
		\end{center}
	\end{figure}

	Depending on  parameter values, the interaction with the environment can strongly localize quantum trajectories by the classical ones~\cite{Spiller1994,Brun1996,Jacobs2000,Pattanayak2018}. 
	Our case is notably different: At any instant of time quantum trajectories are well-localized in the Fock space (Fig.~\ref{fig:1}), 
	but they do not follow the classical mean-field trajectories, as the resulting structure of the probability distribution for the stroboscopic expectation values of $n$ has only 
	a general structural resemblance, see Fig.~\ref{fig:3}(b;top). Nevertheless, working in the essentially quantum regime and tuning  number of bosons $N$, 
	we detect the emergence of the positive largest quantum Lyapunov exponent following the structural chaotization of the asymptotic state, see Fig.~\ref{fig:3}(b). 
	
	It is noteworthy that in the interval $U\in[0.1,0.2]$, where the quantum asymptotic solution undergoes some kind of a `bifurcation', the quantum Lyapunov exponent becomes positive, while the classical mean-field equations still yield a period-1 limit cycle. 
	Stability analysis of the mean-field dynamics gives a clue about possible resolution of this paradox. Indeed, we find that the largest Lyapunov exponent is approaching and almost touches zero line, Fig.~\ref{fig:3}(a), bottom, 
	a signature of bifurcation that is nearly avoided in the mean-field approximation, but is full-fledged in the genuine quantum system. 
	Whether the positive quantum Lyapunov exponent reflects a dynamical property of the system, although the structure of the asymptotic 
	solution lacks an apparent structural complexity, or it reflects a particular quantum-specific effect, is an issue for further studies.

	Finally, we report the result of an extensive numerical experiment aimed at calculating the largest quantum Lyapunov exponent as a function of  the particle interaction strength $U$
	and the amplitude of periodic modulations $\mu_1$. The mean-field system exhibits a variety of regimes on this parameter plane. The quantum phase diagram, in general, more or less follows the classical picture; see Fig.~\ref{fig:4}. However, the quantum case  exhibits a considerably earlier development of chaos and a more complicated structure of regular 
	and chaotic regions. In particular, it follows that multiple two-way transitions are possible, if one of the parameters is 
	fixed and another increased. It can be explained by the reduced complexity of the mean-field model, Eqs.~(\ref{eq:7}). We conjecture 
	that by going to the higher-order mean-field approximations, by increasing the truncation order of accounted correlation functions, and hence,  
	by increasing the dimension of resulting nonlinear system, we would be able to  observe an expansion of the chaotic area. 
	
	\section{Conclusions}
	
	We proposed an approach to calculate the largest  Lyapunov exponent for open quantum systems based on the MCwf  unraveling of the Lindblad equation. 	A numerical realization of this idea allowed us to capture a quantum analogue of the period-doubling route to chaos in a periodically 
	modulated many-body quantum system. 
	The obtained phase diagram on the parameter plane ``interaction strength  -- amplitude of modulations'' revealed a complex structure of regular 
	and chaotic regions with the two-side transitions happening upon the variations of each of the two  parameters. Our findings are relevant to  such fields as quantum electrodynamics, quantum  optics, and
	polaritonic devices, where the quest for the signatures and quantifiers of dissipative quantum chaos is receiving a growing attention. 
	
	Our findings also poses several open issues. 
	The main one is the universality of the largest quantum LE. Namely, there are two interrelated questions: 
	Given a Hamiltonian and dissipators,  what is the proper choice of an  operator to calculate the largest LE? Will its value depend on the operator?  Generally speaking, the answer to the last
	question is `yes'. 
	However, we believe that there must be a certain universality. If two operators are  not too `singular', in the sense that (i) their commutators with the Hamiltonian are 
	not too small and (ii) they are  not too `dark' with respect to the dissipators, then the values of the corresponding LEs will be close to each other.  
	More specifically, a random sampling of the LE operator from the set of all possible (traceless) operators acting in the system Hilbert 
	space, will yield a value close to the average (over the set) value of the largest LE; we plan to corroborate the hypothesis in a more accurate way.
	We also think that this hypothesis can be substantiated by using the concentration-of-measure argument (used, e.g., to prove universality of the microcanonical thermalization \cite{popescu}).
	
	It would be also interesting to check the idea of defining LEs with  out-of-time correlation functions for open systems and compare the corresponding exponents to the one obtained with our approach. With respect to the first part of this program, some steps have already been made \cite{Gal2018_2,Knap2018}.

	The authors acknowledge support of the Russian Foundation for Basic Research grant No.\ 17-32-50078 (IY and SK), Basis Foundation 
	grant No. № 17-12-279-1 (OV, IY and MI), and President of Russian Federation grant No. MD-6653.2018.2 (MI).	Numerical simulations were performed at the Lobachevsky supercomputer (Lobachevsky University, Nizhny Novgorod).

\end{document}